\documentclass{epl}

\title{How short-ranged electrostatics controls the chromatin structure on much
larger scales} \shorttitle{How short-ranged electrostatics etc.}
\author{Helmut Schiessel}
\institute{Max-Planck-Institute for Polymer Research, Theory
Group, POBox 3148, D-55021 Mainz, Germany}

\pacs{87.15.-v}{Biomolecules: structure and physical properties}
\pacs{36.20.Ey}{Conformation (statistics and dynamics}

\begin{document}

\maketitle

\begin{abstract}
We propose that the degree of sweeling of the 30nm chromatin fiber
(a ''measure'' of its transcriptional activity) is mainly
determined by the short-ranged electrostatical interaction between
different sections of the ''folded'' DNA chain. These sections
constitute only a small fraction of the chain and they are located
close to the entry-exit points of the DNA chain at the nucleosome
core particles. We present a model that allows to estimate the
degree of swelling of chromatin fibers as a function of salt
concentration, charge density of the strands etc. Different
mechanisms by which the state of chromatin can be controlled {\it
in vitro} and {\it in vivo} are discussed.
\end{abstract}

DNA in eucaryotic cells is organized within a protein-DNA complex
known as chromatin. In this way plant and animal genomes are
compactified into volumes whose linear dimensions are many orders
of magnitude smaller than their contour lengths. For instance, the
human genom is made up of billions of base pairs (bp)
corresponding to about one meter of DNA chains. These
highly-charged and hard-to-bend polymers are condensed into
complexes that have a characteristic size of a micron and
therefore fit into the cell nucleus. {\it At the same time} it is
of vital importance that a fraction of the genetic code stored
within these tight complexes is accessible: For instance, gene
regulatory proteins bind to specific sequences, the rather bulky
protein complex RNA polymerase needs to gain access to a whole
gene during its transcription etc. How DNA is ''folded'' within
chromatin and how it can be ''unfolded'' for, say, transcription
purposes is still poorly understood.

The primary structure of chromatin is known in great detail from
x-ray studies~\cite{luger97}. The basic unit is the nucleosome
consisting of the core particle and the linker DNA (of typical
length 60bp) that connects to the neighboring core particle. The
resulting structure is a beads-on-chain necklace (10nm fiber). The
core particle consists of 147bp DNA wrapped in 1- and $3/4$
left-helical turns around a globular octamer of cationic proteins
(core histones) -- forming a squat cylinder with a radius of about
5{\it nm} and height of about 6{\it nm}. The higher-order
secondary and tertiary structures on scales from 10{\it nm} up to
a micron are still a matter of controversy~\cite{woodcock01}. It
is well-known that the bead-on-a-chain folds into a thicker fiber
with a diameter of roughly 30{\it nm} but it is still not clear
how the neighboring beads are arranged in this fiber with respect
to each other and where the linker is located. In the solenoid
models~\cite{finch76} it is assumed that the chain of nucleosomes
forms a helical structure with the linker DNA being bent whereas
the zig-zag- or crossed-linker models~\cite{woodcock93} posits
straight linkers that connect core particles that are located at
opposite sites of the fiber. An example of that kind of structure,
the two-dimensional zig-zag fiber, is shown in Fig.~1(a). In
general, the fiber is three-dimensional.

The difficulties in determining the structure of the 30nm fibers
are due to the lack of reliable experimental methods. Electron
cryomicroscopy allows to visulize fibers {\it in vitro} but the
fiber geometry can only be detected at ionic strengths well beyond
physiological conditions where the structures are much more
open~\cite{bednar98}. Fibers at such low ionic strengths show
straight linkers, a fact that supports the second class of the
above mentioned chromatin models -- at least for fibers at these
unphysiological conditions. At higher ionic strengths, however,
the fiber is so dense that their internal structure remains
obscure.

A different approach allows to obtain structural information via
an indirect method: the stretching of chromatin fibers with the
help of micromanipulation devices as it was achieved recently. Cui
and Bustamante~\cite{cui2000} found that chromatin fibers are
extremely soft with respect to their extension (compared to naked
DNA). In principle, it should be possible to compare the obtained
experimental data (namely the force-extension curves) with the
existing models and to determine which is the appropriate model.
In the meantime the crossed-linker model was indeed reevaluated in
view of the stretching experiment. Katritch {\it et
al}.~\cite{katritch00} performed computer simulations showing that
this model is capable of reproducing experimental data quite well
for several sets of values of the ''free'' parameters. Schiessel
{\it et al}.~\cite{schiessel01} gave an analytical treatment and
introduced optimization criteria (see below) that led to
reasonable predictions of the mechanical properties of the
chromatin fiber {\it without the use of any adjustable
parameters}. These studies do not invalidate the solenoid model
but demonstrate at least that the crossed-linker model is
successful in describing some characteristic features of the 30nm
fiber.

The main idea of the crossed-linker model as it was introduced by
Woodcock and coworkers~\cite{woodcock93} is that the
three-dimensional structure of the 30nm chromatin fiber is
determined by two quantities only: the entry-exit angle $\alpha $
of the linker DNA at the core particle ({\it cf.} Fig.~1(a)) and
the linker length $b$. The entry-exit angle $\alpha $ is
determined by local properties of the nucleosome core particle and
it seems reasonable to assume that this quantity is constant
throughout the fiber as long as there are homogeneous conditions.
The second quantity, the linker length $b$, determines the
rotational setting of neighboring nucleosomes. This follows from
the fact that there are {\it specific} binding sites on the
histone octamer where the DNA adsorbes with its minor groove
facing
inwards. Taking into account that the helical repeat length is $\approx 10$%
bp it is clear that increasing $b$ by one bp leads to rotation of
a nucleosome by $\approx 36{{}^{\circ }}$ with respect to its
neighbor. As a consequence, changing $b$ by multiples of $\approx
10$bp does not affect the rotational setting, i.e., the dihedral
angle $\phi $ is periodic function of the linker length.

It was found that chromatin fibers have the tendency to show a
preferential {\it quantization} with respect of their linker
length and that this quantization is set by the helical
twist~\cite{widom92}. This indicates that the dihedral angle $\phi
$ shows a preferential value and thus the relative rotational
setting is more or less constant throughout the fiber. Therefore
it seems to be a reasonable approximation to consider both, $\alpha $ and $%
\phi $ to be constant throughout the fiber (together with $b$).
The geometrical properties of the fiber, {\it i.e.}, its radius
$R$ and the linear density $\Lambda$
of nucleosomes along the fiber, are then a function of the two angles $%
\alpha $ and $\phi $ only (and of $b$ in a trivial sense). In Ref.
\cite {schiessel01} we termed this geometrical model therefore the
{\it two-angle model}. A specific example of such a two-angle
fiber with $\phi =180$ is the zig-zag fiber that we depicted in
Fig. 1(a).

\begin{figure}
\onefigure{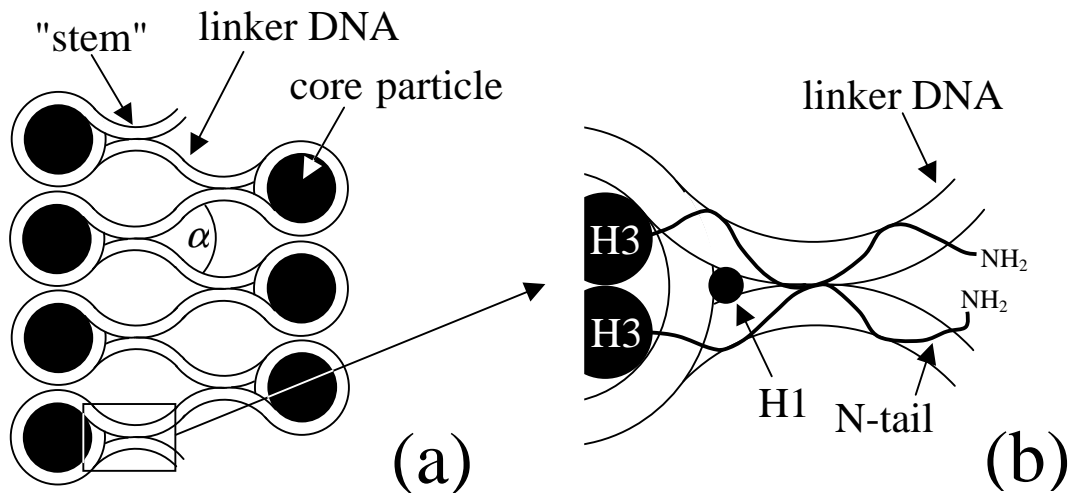} \caption{(a) Schematic view of a section of
the 30nm chromatin fiber. For simplicity, the fiber is shown as a
two-dimensional zig-zag fiber. The histone octamers are displayed
as black circular disks. The DNA (white) consists of wrapped
parts, ''stem'' sections and sections that link neighboring
nucleosomes. (b) Enlarged view of the stem region. As shown
schematically the stem section of the DNA interacts with cationic
N-tails of the core histones and with the linker histone H1 (see
text for details).}
\end{figure}

In our study of the geometrical and mechanical properties of the
two-angle model~\cite{schiessel01} we suggested that the values of
$\alpha $ and $\phi $ of the ''native'' 30nm chromatin fiber in
its ''silenced'' form (no transcriptional activity) are chosen in
such a way that the
three-dimensional {\it density} of the fiber (defined by $\Lambda /\pi R^{2}$%
) and its ''{\it accessibility''} are optimized. Accessibility
means here how rapid the ''silenced'' fiber opens up when its
geometrical properties (characterized by the two angles $\phi $
and $\alpha $) are changed for the purpose of, say, transcription.
As long as the positions of the nucleosomes at the DNA are fixed
(as assumed here) $\phi $ and $b$ are always constant and hence it
is the angle $\alpha $ with which the fiber geometry, i.e., its
degree of swelling, is controlled. High accessibility means that
the fiber opens up strongly with an increasing value of $\alpha$,
leading to a strong reduction of the nucleosome line density.
Hence our measure for accessibility, as introduced in
Ref.~\cite{schiessel01}, is the quantity $-\partial
\Lambda/\partial \alpha $.

The purpose of this paper is to study how the entry-exit angle of
the DNA at the nucleoseomes can be controlled via electrostatics.
It can been seen in the cryo EM studies~\cite{bednar98} that the
fibers open up and become therefore more accessible when the ionic
strength is reduced and that this opening is directly linked to an
increase in $\alpha $. It was suggested that via other mechanisms
(for instance, the acetylation of so-called histone
tails~\cite{vanholde96}, as explained in more detail below) the
angle $\alpha $ and therefore the degree of swelling can be
changed for a given section of the fiber and that this constitutes
a biochemical means to control the transcriptional activity of
genes.

Whereas the x-ray studies of the core particle~\cite{luger97}
allow a detailed knowledge of the wrapped part of DNA it does not
give insight into the conformational properties of the entering
and exiting strands since the core particles were constituted from
146bp DNA only. One has therefore to refer to the electron
cryomicrographs. In these micrographs it can be seen clearly that
10nm stretches of the entering and exiting DNA strands are glued
together forming a unique ''stem motif''~\cite{bednar98} ({\it
cf.} also Fig.~1(a)). The glueing of the two equally charged
chains is accomplished -- amongst other things -- via a special
cationic protein, the so-called linker histone H1 as shown
schematically in Fig. 1(b).

At physiological concentrations (roughly 100mM salt corresponding
to a screening length of about 10\AA ) the electrostatics is
essentially short-ranged (note that the diameter of DNA is 20\AA
). It seems therefore reasonable to assume that $\alpha $ is set
within the small region where the two linker DNA are in close
contact, i.e., in the stem region~\cite{remark0}. This value of
$\alpha $ in turn controls the large-scale secondary structure of
chromatin, the 30nm-fiber. To mimic this situation we assume in
the following two parallel DNA strands that are hold together
tightly at $y=0$ for $x\leq 0$ and that are free for $x>0$, cf.
Fig.~2. Because of their mutual electrostatic repulsion the two
strands bend away from each other. When the two strands are far
enough from each other their interaction is screened so that they
asymptotically approach straight lines (neglecting thermal
fluctuations as it is appropriate for the length scales under
consideration). The two asymptotes define the opening angle
$\alpha $ as indicated in Fig.~2. It is the purpose of this study
to determine this angle and its dependence on different physical
parameters~\cite{remark1}.

\begin{figure}
\onefigure{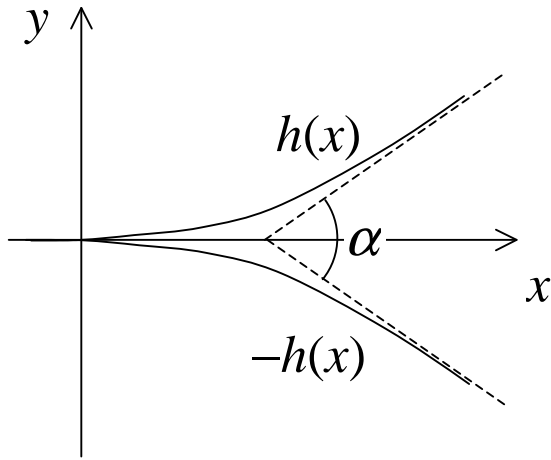} \caption{Idealized model for the entry-exit
region of the DNA at the nucleosome. $h\left( x\right) $ and
$-h\left( x\right) $ describe the conformations of the entering
and exiting DNA sections, respectively. The asymptotic slope at
large $x$ determines the entry-exit angle $\alpha $ and, in turn,
the overall conformation of the chromatin fiber (as depicted in
Fig. 1(a)).}
\end{figure}

We describe the conformation of the upper DNA chain by the height function $%
h\left( x\right) $. By symmetry the position of the lower strand
is then given by $-h\left( x\right) $. From the above given
considerations follow two boundary conditions at $x=0$:
\begin{equation}
h\left( 0\right) =h^{\prime }\left( 0\right) =0  \label{bound1}
\end{equation}
The entry-exit angle $\alpha $ is related to the slope of $h\left(
x\right) $ at infinity as follows:
\begin{equation}
\tan \left( \alpha /2\right) =h^{\prime }\left( \infty \right)
\label{alpha}
\end{equation}
We model the two DNA chains as semiflexible polymers with
persistence length $l_{P}$ and line-charge density $\lambda $
immersed in a salt solution characterized by the Bjerrum length
$l_{B}\equiv e^{2}/\epsilon k_{B}T$ and the Debye screening length
$\kappa ^{-1}=\left( 8\pi c_{s}l_{B}\right)
^{-1/2}$ ($\epsilon $ dielectric constant of the solvent, $T$ temperature, $%
k_{B}$ Boltzmann constant and $c_{s}$ salt concentration). We
assume that the charges on the DNA strands interact via a screened
electrostatic potential. The free energy of the system is then
given by
\begin{equation}
\frac{F}{k_{B}T}\simeq \int_{0}^{\infty }dx\left[ l_{P}\left( \frac{d^{2}h}{%
dx^{2}}\right) ^{2}+2l_{B}\lambda ^{2}K_{0}\left( 2\kappa h\left(
x\right) \right) \right]  \label{freee}
\end{equation}
The first term in the integral accounts for the bending of the
{\it two} DNA
strands and the second one describes the interaction between the two chains (%
$K_{0}\left( x\right) $ being the $0th$ order modified Bessel
function). Here the interaction of a given charge on one chain
with all the charges on the other chain is approximated by the
interaction of this charge with a {\it straight} chain at the
distance $2h$. The integral, eq.~\ref{freee}, is a good
approximation as long as the value of $\alpha $ that follows from
its minimization is sufficiently small, i.e., as long as
$h^{\prime }\left( x\right) \ll 1$ for all $x$. We furthermore
neglect here the interaction
between charges on the same chain. The conformation of the upper chain, $%
h\left( x\right) $, is then the solution of the corresponding
Euler-Lagrange equation (using $K_{0}^{\prime }\left( x\right)
=-K_{1}\left( x\right) $)
\begin{equation}
l_{P}\frac{d^{4}h}{dx^{4}}-2l_{B}\lambda ^{2}\kappa K_{1}\left(
2\kappa h\right) =0  \label{euler1}
\end{equation}
together with four boundary conditions: Two are given at the
origin, eq.~\ref {bound1}, and two follow from the condition of
straight ''linkers'' at infinity:
\begin{equation}
h^{\prime \prime }\left( \infty \right) =h^{\prime \prime \prime
}\left( \infty \right) =0  \label{bound2}
\end{equation}
Defining $\tilde{h}=2\kappa h$ and introducing the dimensionless quantity $%
\tilde{x}=\left( 4l_{B}\lambda ^{2}\kappa ^{2}l_{P}^{-1}\right)
^{1/4}x$, eq.~\ref{euler1} can be rewritten as
\begin{equation}
\frac{d^{4}\tilde{h}}{d\tilde{x}^{4}}-K_{1}\left( \tilde{h}\right)
=0 \label{dgl}
\end{equation}
We denote the dimensionless asymptotic slope by $c_{0}=\left. d\tilde{h}/d%
\tilde{x}\right| _{\tilde{x}=\infty }$. It follows immediately
that $\tan \left( \alpha /2\right) $ is given by
\begin{equation}
\tan \left( \alpha /2\right) =h^{\prime }\left( \infty \right) =\frac{c_{0}}{%
\sqrt{2}}\left( \frac{l_{B}}{l_{P}}\right) ^{1/4}\left( \frac{\lambda }{%
\kappa }\right) ^{1/2}  \label{tan}
\end{equation}

Equation~\ref{dgl} can be solved asymptotically for small value of
$\tilde{h}$ where $K_{1}\left( \tilde{h}\right) \simeq
1/\tilde{h}$. In leading order the asymptotic solution is given by
$\tilde{h}\left( \tilde{x}\right) \simeq
\tilde{h}_{0}\left( \tilde{x}\right) =\tilde{x}^{2}\sqrt{\left( -\ln \tilde{x%
}\right) }$. This solution fulfills the boundary conditions at the
origin, eq.~\ref{bound1}. The boundary conditions at infinity,
eq.~\ref{bound2}, are only marginally affected by the solution for
small values of $\tilde{x}$ as can be seen by adding terms like
$a\tilde{x}^{2}$ and $b\tilde{x}^{3}$. We
can use this approximate solution to give a rough estimate for $c_{0}$. $%
\tilde{h}_{0}\left( \tilde{x}\right) $ reaches its maximal slope $%
0.\,\allowbreak 56$\ at $\tilde{x}\simeq 0\allowbreak
.\,\allowbreak 41$.
For larger values of $\tilde{x}$ the approximate solution $\tilde{h}%
_{0}\left( \tilde{x}\right) $ becomes more and more unreliable.
This
asymptotic analysis indicates that $c_{0}$ is of the order one\cite{remark2}%
. The asymptotic solution allows to estimate the free energy of
the configuration since most of the bending and electrostatic
contributions are localized close to the origin. Replacing
$h\left( x\right) $ in eq.~\ref {freee} by $h_{0}\left( x\right)
=\left( \gamma x\right) ^{2}\sqrt{-\ln \left( \gamma x\right)
}/2\kappa $ with $\gamma =\left( 4l_{B}\lambda ^{2}\kappa
^{2}/l_{P}\right) ^{1/4}$ and integrating from $x=0$ to $x\approx
\gamma ^{-1}$ (neglecting logarithmic contributions) we find that
both, the
bending contribution and the electrostatic contribution scale as $%
F/k_{B}T\approx l_{B}^{3/4}\lambda ^{3/2}l_{P}^{1/4}\kappa
^{-1/2}$.

Our calculation does not account for {\it intra}molecular
electrostatic contributions. These can be included by adding the
electrostatic persistence length $l_{OSF}=l_{B}\lambda
^{2}/4\kappa ^{2}$ (the Odijk-Skolnick-Fixman length, cf.
Ref.~\cite{barrat96}) to the bare persistence length, i.e., by
replacing $l_{P}$ by $l_{P}+l_{OSF}$. This leads to $\tan \left(
\alpha
/2\right) \propto \left( l_{OSF}/\left( l_{P}+l_{OSF}\right) \right) ^{1/4}$%
. Note that for small salt concentrations (large $\kappa ^{-1}$)
$l_{OSF}$ becomes increasingly important so that the angle
approaches a universal value independent of $\kappa ^{-1}$,
$\lambda $ and $l_{B}$ -- at least as long as the above stated
approximations are still applicable in this limit. For the salt
concentrations under consideration $l_{OSF}\ll l_{P}$ so that we
disregard intramolecular contributions in the following.

We apply now our results to the problem of how the geometry of the
chromatin fiber is controlled {\it in vitro} and, on a more
tentative level, {\it in vivo}. The {\it in vitro}-experiments
show that chromatin fibers ''open up'' with a decreasing salt
concentration. From the electron cryomicrographs it was found that
$\alpha \approx 85{{}^{\circ }}$ for $c_{s}=5mM$ and $\alpha
\approx 45{{}^{\circ }}$ for $c_{s}=15mM$~\cite{bednar98}.
Furthermore, from electron cryotomography it was estimated that
$\alpha \approx 35{{}^{\circ }} $ for
$c_{s}=80mM$~\cite{bednar98}. We expect from eq.~\ref{tan} that
$\alpha \simeq 2\arctan \left( Cc_{s}^{-1/4}\right) $ with $C$
being a constant. Let us take the angle at the highest salt
concentration, $c_{s}=80mM$, as the reference value. From this
follows $C=0.94$. With this value of $C$ we predict $\alpha
\approx 51{{}^{\circ }}$ for $c_{s}=15mM$ and $\alpha
\approx 64{{}^{\circ }}$ for $c_{s}=80mM$. Whereas the predicted value of $%
\alpha $ at the intermediate ionic strength is close to the
experimental one, the value for low salt concentrations is
noticably too low. However, we expect the theory to break down for
such large values of $\alpha $.

How can the degree of swelling of the chromatin fiber be
controlled {\it in vivo}? Under the assumption that the above
mentioned geometry is valid the only parameter that might be under
biochemical control is the linear charge density $\lambda $. It is
known that there are flexible cationic tails, namely the C-tail of
the linker histone as well as the lysine-rich N-tails of the core
histones involved in determining the degree of swelling. In Fig.
1(b) we give a tentative picture of the conformation of two
N-tails that protrude from the histone core (we assume here that
these are the tails of the two so-called H3 core histones that are
located close to the entry-exit point). It is known that if either
of these components is missing the fiber does not fold
appropriately (cf. Ref.~\cite{vanholde96} and references therein).
As indicated in the Figure the tails might form a complex with the
entering and exiting linker DNA in such a way that they
effectively reduce its linear charge density $\lambda $. It is
known that transcriptionally active regions in chromatin show an
acetylation of the core histone tails (i.e., the cationic groups
of the lysines are neutralized). In our tentative picture this
acetylation mechanism would increase $\lambda $ and according
to eq.~\ref{tan} this would lead to an opening of the entry-exit angle $%
\alpha $. The acetylation might therefore be the first step in the
decondensation of a stretch of the chromatin fiber that needs to
be accessed for transcription~\cite{remark3}.

Further steps might then involve the loss of the linker histones leading to $%
\alpha \approx \pi $, i.e., a bead-on-string filament. After the
loss of the linker histones the nucleosomes might even become
mobile as it was reported in Ref.~\cite{pennings94}. We recently
suggested that this ''nucleosome sliding'' results from DNA
''reptating'' around the histone core with the help of
intranucleosomal loops~\cite{schiessel01b}. Loop formation might
also be crucial for the actual transcription itself: It was
suggested that RNA polymerase might elongate through the
nucleosome by passing it in a loop~\cite{studitsky94}. However, up
to now a detailed knowledge of how transcription through chromatin
is possible is still unclear and a matter of current research.

\acknowledgments I wish to thank R. Bruinsma, A. Johner and G.
Migliorini for useful discussions.


\begin{thebibliography}{0}

\bibitem{luger97}
  \Name{Luger K., M\"{a}der A. W., Richmond R. K., Sargent D. F.
\and T. J. Richmond}
  \REVIEW{Nature}{389}{1997}{251}.

\bibitem{woodcock01}
  \Name{Woodcock C. L. \and Dimitrov S.}
  \REVIEW{Curr. Opim. Genet. Dev.}{11}{2001}{130}.

\bibitem{finch76}
  \Name{Finch J. T. \and Klug A.}
  \REVIEW{Proc. Natl. Acad. Sci. USA}{73}{1976}{1897}.

\bibitem{woodcock93}
  \Name{Woodcock C. L., Grigoryev S. A., Horowitz R. A., \and Whitaker N.}
  \REVIEW{Proc. Natl. Acad. Sci. USA}{90}{1993}{9021}.

\bibitem{bednar98}
  \Name{Bednar J., Horowitz R. A., Grigoryev S. A.,
Carruthers L. M., Hansen J. C., Koster A. J. \and Woodcock C. L.}
  \REVIEW{Proc. Natl. Acad. Sci. USA}{95}{1998}{14173}.

\bibitem{cui2000}
  \Name{Cui Y. \and Bustamante C.}
  \REVIEW{Proc. Natl. Acad. Sci. USA}{97}{2000}{127}.

\bibitem{katritch00}
  \Name{Katritch V., Bustamante C. \and Olson W. K.}
  \REVIEW{J. Mol. Biol.}{295}{2000}{29}.

\bibitem{schiessel01}
  \Name{Schiessel H., Gelbart W. M. \and Bruinsma R.}
  \REVIEW{Biophys. J.}{80}{2001}{1940}.

\bibitem{widom92}
  \Name{Widom J.}
  \REVIEW{Proc. Natl. Acad. Sci. USA}{89}{1992}{1095}.

\bibitem{vanholde96}
  \Name{van Holde K. \and Zlatanova J.}
  \REVIEW{Proc. Natl. Acad. Sci. USA}{93}{1996}{10548}.

\bibitem{remark0}
We assume here that internucleosomal interaction can be neglected.
There is evidence from the stretching experiments\cite{cui2000}
that the nucleosomes are ''in contact'' at a physiological ionic
strength and that there is a short-range attraction between them.
With the swelling of the fiber at lower ionic strength the
internucleosomal interaction becomes less important and might be
negligible (cf. our discussion of the internucleosomal interaction
in Ref. \cite{schiessel01}).

\bibitem{remark1}
A similar problem is that of a charged chain that is deflected by
a rod perpendicular to it. This mimics a situation occuring in
semidilute polyelectrolyte solutions and is discussed in Ref.
\cite{barrat96}, p 39.

\bibitem{barrat96}
  \Name{Barrat J.-L. \and Joanny J.-F.}
  \REVIEW{Adv. Chem. Phys.}{94}{1996}{1}.

\bibitem{remark2}  Note that our analysis is based on the assumption of a
vanishing distance between the two strands for $x\leq 0$. This
leads to a logarithmic singularity of the curvature at $x=0$,
i.e., to an infinite torque at the origin. The structure could
only be stabilized for an infinite adsorption energy per length
between the two strands for $x\leq 0$. A more
realistic case is $\tilde{h}\left( x\right) \equiv \varepsilon \ll 0$ for $%
x\leq 0$. In this case the following approximate solution can be
constructed: $\tilde{x}^{4}/24\varepsilon -\tilde{x}^{3}/\left( 3\tilde{x}%
_{0}\sqrt{-\ln \tilde{x}_{0}}\right) +\tilde{x}^{2}\sqrt{\left( -\ln \tilde{x%
}_{0}\right) }+\varepsilon $ for $\tilde{x}<\tilde{x}_{0}$ and $\tilde{h}%
_{0}\left( x\right) +\varepsilon $ for $\tilde{x}_{0}<\tilde{x}\ll 1$; here $%
\tilde{x}_{0}$ is the solution of $\tilde{x}_{0}^{2}\sqrt{-\ln \tilde{x}_{0}}%
=\varepsilon $. These two functions cross over smoothly at
$\tilde{x}\approx \tilde{x}_{0}$ and obey Eq. \ref{dgl} for
$\tilde{x}\ll \tilde{x}_{0}$ and for $\tilde{x}\gg \tilde{x}_{0}$,
respectively. This indicates that the solution is only marginally
affected by the value of $\varepsilon $ as long as $\varepsilon
\ll 0$ and that our original assumption, Eq. \ref{bound1}, is
reasonable.

\bibitem{remark3}  The processes that are involved in the acetylation and
deacetylation might be quite specific and involved as, for
instance, discussed in Ref. \cite{grunstein97}. The histone tail
modifications might serve specific functions via the modification
of their secondary structure
that in turn modifies their interaction with certain proteins\cite{hansen98}%
. Recently there are even attempts to decipher a specific
''language'' of covalent histone modifications\cite{strahl00}. It
might be that such specific processes act in concert with the more
basic charge neutralization principle discussed here.

\bibitem{grunstein97}
  \Name{Grunstein M.}
  \REVIEW{Nature}{389}{1997}{349}.

\bibitem{hansen98}
  \Name{Hansen J. C., Tse C. \and Wolffe A. P.}
  \REVIEW{Biochemistry}{37}{1998}{17637}.

\bibitem{strahl00}
  \Name{Strahl B. D. \and Allis D.}
  \REVIEW{Nature}{403}{2000}{41}.

\bibitem{pennings94}
  \Name{Pennings S., Meersseman G. \and Bradbury E. M.}
  \REVIEW{Proc. Natl. Acad. Sci. USA}{91}{1994}{10275}.

\bibitem{schiessel01b}
  \Name{Schiessel H., Widom J., Bruinsma R. F. \and Gelbart W. M.}
  \REVIEW{Phys. Rev. Lett.}{86}{2001}{4414}.

\bibitem{studitsky94}
  \Name{Studitsky V. M., Clark D. J. \and Felsenfeld G.}
  \REVIEW{Cell}{71}{1992}{371}.

\end{thebibliography}
\end{document}